# Effect of crystalline anisotropy on vertical ($\bar{2}$01) and (010) *β*-Ga$_2$O$_3$ Schottky barrier diodes on EFG single-crystal substrates

Houqiang Fu, Hong Chen, Xuanqi Huang, Izak Baranowski, Jossue Montes, Tsung-Han Yang, and Yuji Zhao*

*School of Electrical, Computer and Energy Engineering, Arizona State University, Tempe, AZ 85287, U.S.A.*

*Email address: yuji.zhao@asu.edu

*Abstract*

Vertical ($\bar{2}$01) and (010) *β*-Ga$_2$O$_3$ Schottky barrier diodes (SBDs) were fabricated on single-crystal substrates grown by edge-defined film-fed growth (EFG) method. High resolution X-ray diffraction (HRXRD) and atomic force microscopy (AFM) confirmed good crystal quality and surface morphology of the substrates. The electrical properties of both devices, including the current-voltage (I-V) and capacitance-voltage (C-V) characteristics, were comprehensively measured and compared. The ($\bar{2}$01) and (010) SBDs exhibited on-resistances ($R_{on}$) of 0.56 and 0.77 mΩ·cm$^2$, turn-on voltages ($V_{on}$) of 1.0 and 1.3 V, Schottky barrier heights (SBH) of 1.05 and 1.20 eV, electron mobilities of 125 and 65 cm$^2$/(V·s), respectively, with a high on-current of ~ 1.3 kA/cm$^2$ and on/off ratio of ~10$^9$. The (010) SBD had a larger $V_{on}$ and SBH due to anisotropic surface properties (i.e., surface Fermi level pinning and band bending), as supported by X-ray photoelectron spectroscopy (XPS) measurements. Temperature-dependent I-V also revealed the inhomogeneous nature of the SBH in both devices, where ($\bar{2}$01) SBD showed a more uniform SBH distribution. The homogeneous SBH was also extracted: 1.33 eV for ($\bar{2}$01) SBD and 1.53 eV for (010) SBD. The reverse leakage current of the devices was well described by the two-step trap-assisted tunneling model and the one-dimensional variable range hopping conduction (1D-VRH) model. The ($\bar{2}$01) SBD showed a larger leakage current due to its lower SBH and smaller activation energy. These results indicate the crystalline anisotropy of *β*-Ga$_2$O$_3$ can affect the electrical properties of vertical SBDs and should be taken into consideration when designing *β*-Ga$_2$O$_3$ electronics.

Beta-phase gallium oxide ($\beta$-Ga$_2$O$_3$) is a newly emerged candidate in the wide bandgap (WBG) semiconductor family, which has attracted considerable attention for efficient power conversion applications[1-3] in smart grids, renewable energy, data center power supply, and automotive electronics. Compared with other WBG semiconductors, such as SiC and GaN, $\beta$-Ga$_2$O$_3$ has larger bandgap $E_g$ (4.8 eV) and breakdown electric field $E_{br}$ (~ 8 MV/cm).[4-6] To evaluate the performance of power electronics, the Baliga's figure of merit (FOM) ($\varepsilon_r \mu E_{br}^3$ where $\varepsilon_r$ is the relative dielectric constant and $\mu$ is the mobility), is often used.[6] $\beta$-Ga$_2$O$_3$ exhibits 4 times larger Baliga's FOM than SiC and GaN,[5,6] indicating $\beta$-Ga$_2$O$_3$ power electronics have the potential to outperform SiC and GaN devices.

One of the important advantages of $\beta$-Ga$_2$O$_3$ is the availability of cost-effective single-crystal substrates.[1-6] Edge-defined film-fed growth (EFG) is one of the most popular methods for producing large-sized $\beta$-Ga$_2$O$_3$ substrates due to its low cost and compatibility with mass production.[6-8] High quality two inch ($\bar{2}$01) substrates grown by the EFG method have been commercialized with controllable doping concentrations ranging from $10^{16}$ cm$^{-3}$ to $10^{19}$ cm$^{-3}$ and four inch substrates have also been demonstrated.[7,9] Substrates on other crystal orientations such as (010),[3] (001),[10] and (100)[6] can also be grown by EFG method. Electronic devices on various crystal orientations have already been reported, such as field-effect transistors (FETs)[1,2] and Schottky barrier diodes (SBDs)[3,6,7,10].

Because $\beta$-Ga$_2$O$_3$ has a highly asymmetric monoclinic crystal structure, the anisotropic material properties of $\beta$-Ga$_2$O$_3$ have garnered tremendous interest.[11-14] For example, Guo et al.[11] found a large anisotropy in the thermal properties of $\beta$-Ga$_2$O$_3$ where [010] direction had a ~3 times larger thermal conductivity than [100] direction. Chen et al.[12] observed anisotropic nonlinear optical properties such as the two-photon absorption coefficient and the Kerr refractive index on ($\bar{2}$01) and (010) $\beta$-Ga$_2$O$_3$. Wong et al.[13] reported electron mobility anisotropy in $\beta$-Ga$_2$O$_3$ FETs due to the anisotropic carrier scattering caused by asymmetric phonon modes.[14] In addition, evidence has also shown surface properties of $\beta$-Ga$_2$O$_3$ are also anisotropic due to different atomic configurations and dangling bonds on different orientations.[15-17] For example, Sasaki et al.[15] found the growth rate of (100) $\beta$-Ga$_2$O$_3$ was significantly lower than other orientations due to the low adhesion energy on the terraces of the (100) surface. Hogan et al.[16] studied the dry etching of $\beta$-Ga$_2$O$_3$ by inductively coupled plasma (ICP) and obtained much lower etching rate on (100) than on (010) and ($\bar{2}$01) due to the surface oxygen anions and lower dangling bond density. Similar anisotropic etch rates were also observed in the KOH wet etching of β-Ga$_2$O$_3$.[17]

From a device perspective, these anisotropic bulk and surface material properties of β-Ga$_2$O$_3$ are expected to have an impact on the performances of β-Ga$_2$O$_3$ electronic devices. Similar phenomena have already been observed in wurtzite III-nitride semiconductors, where devices grown on polar, semipolar and nonpolar orientations have shown distinct behaviors.[18-21] However, systematic study on the effect of crystalline anisotropy on the β-Ga$_2$O$_3$ electronic devices is still lacking. In this work, we fabricated vertical ($\bar{2}$01) and (010) β-Ga$_2$O$_3$ SBDs on EFG single-crystal substrates and comprehensively compared their electrical properties. The crystal orientations and associated surface properties do impact the behaviors of the SBDs at both forward and reverse bias, in terms of turn-on voltage ($V_{on}$), Schottky barrier height (SBH), electron mobility, and reverse leakage current.

The β-Ga$_2$O$_3$ single-crystal substrates were grown by EFG method. High purity Ga$_2$O$_3$ powder was the source material and tin oxide (SnO$_2$) powder was the precursor for n-type Sn dopants. The powder mixture was melted by radio-frequency heating, after which a β-Ga$_2$O$_3$ seed crystal was used to initialize the crystal growth. More information about growth methods can be found elsewhere.[8] The crystal quality of the substrates was characterized by PANalytical X'Pert Pro high resolution X-ray diffraction (HRXRD) using Cu K$\alpha$1 radiation with a wavelength of 1.541 Å. Hybrid monochromator and triple axis module were used as incident and diffracted beam optics, respectively. Figures 1(a) and 1(b) present the rocking curves (RCs) for the ($\bar{2}$01) and (010) β-Ga$_2$O$_3$ substrates, respectively. The full width at half maximum (FWHM) of the ($\bar{2}$01) RC was 87 arc sec and the FWHM of the (020) RC was 90 arc sec. This indicates both substrates have similar crystal quality with a dislocation density in the high $10^6$ cm$^{-2}$ range based on the methods described in Ref. 22. Bruker's Multimode atomic force microscopy (AFM) was used to examine the surface morphology of the substrates and the representative images were shown in Figs. 1(c) and 1(d). The root-mean-square (RMS) roughness of the $4 \times 4$ µm$^2$ scanning area of the substrates was 0.1-0.2 nm. HRXRD and AFM results indicate high quality β-Ga$_2$O$_3$ substrates with low dislocation density and good surface morphology were obtained.

Figure 2 shows the schematic unit cell of β-Ga$_2$O$_3$ and atomic configurations of ($\bar{2}$01) and (010) planes. β-Ga$_2$O$_3$ crystallizes into a monoclinic structure (C2/m) with lattice constants $a$ = 1.223 nm, $b$ = 0.304 nm, and $c$ = 0.580 nm and angles $\alpha = \gamma = 90°$, and $\beta = 104°$.[16,17] There are two gallium sites: tetrahedrally-coordinated $Ga_I$ (4 bonds) and octahedrally-coordinated $Ga_{II}$ (6 bonds), and three oxygen sites: $O_I$ (3 bonds), $O_{II}$ (3 bonds) and $O_{III}$ (four bonds). ($\bar{2}$01) and (010) surfaces differ significantly in atomic configurations and dangling bonds. The ($\bar{2}$01) surface is exclusively terminated either by $Ga_I$ or $Ga_{II}$ or oxygen, while the (010) surface is composed of both gallium ($Ga_I$ and $Ga_{II}$) and oxygen with a Ga-to-O ratio of 2:3.[16,17] To assess surface properties of the ($\bar{2}$01) and (010) substrates, X-ray photoelectron spectroscopy (XPS) measurements were carried out using a monochromated Al K$\alpha$ X-ray source under an ultrahigh vacuum of <10$^{-9}$ Torr. The system was calibrated by the standard reference C 1s peak. The valance band minimum ($E_{VBM}$) can be extracted by linearly extrapolating the leading edge of the valance band (VB) spectra to the baseline, as shown in Fig. 3. The procedure of calculating surface band bending of semiconductors is detailed in Ref. 23. In n-type semiconductors, due to surface states and defects, the Fermi level is pinned at the charge neutrality level (CNL) at the surface, where the surface barrier height $\Phi_{surf}$ is given by[23]

$$\Phi_{surf} = E_g - E_{VBM} \quad (1)$$

The obtained $\Phi_{surf}$ was 1.14 eV for the ($\bar{2}$01) surface and 1.63 eV for the (010) surface. The conduction bands (CBs) were bent upward, indicating the presence of negatively charged surface states and defects at the surfaces.[23] This is partly responsible for the difficulty in forming ohmic contacts to $\beta$-$Ga_2O_3$.[1,6] The (010) surface has a 0.49 eV larger band bending, which explains the fact that it is more difficult to realize ohmic contacts on the (010) orientation.[1,2,17,24] The different crystal structures and surface properties of ($\bar{2}$01) and (010) $\beta$-$Ga_2O_3$ could impact the electrical properties of devices based on them. This following section will use $\beta$-$Ga_2O_3$ vertical SBDs as a case study and comprehensively compare their device performances.

The SBDs were fabricated on the ($\bar{2}$01) and (010) $\beta$-$Ga_2O_3$ substrates using optical photolithography. The two substrates were cleaned in acetone and isopropyl alcohol. The backside surfaces of the substrates were treated by the $BCl_3$-based ICP etching for 5 minutes at an ICP source/bias power of 400/30 W, a $BCl_3$/Ar flow rate of 20/5 sccm and a pressure of 15 mTorr.[16] The etch rate was ~ 20 nm/min and the total etching thickness was ~ 100 nm. The etching process can create donor-like surface defects to facilitate ohmic contacts.[1,6] For ohmic contacts, Ti/Al/Ti/Au metal stacks were formed on the backside of the substrates using electron beam evaporation, followed by rapid thermal annealing (RTA) at 470 °C in nitrogen for 1 minute. For the circular Schottky contacts (diameter of 100 µm), Pt/Au metal stacks were deposited by electron beam evaporation. The electrical measurements were carried out on a probe station with a controllable thermal chuck using Keithley 2410 sourcemeter and 4200-SCS parameter analyzer.

Figure 4(a) shows the forward I–V characteristics of the ($\bar{2}$01) and (010) SBDs at room temperature (RT) in linear scale. The measurement apparatus has a upper current limit of 0.1 A. The forward current of the devices exceeded 0.1 A at a voltage of 1.7 V in the ($\bar{2}$01) SBD and 2.4 V in the (010) SBD. The $V_{on}$ of ($\bar{2}$01) and (010) SBDs were 1.0 V and 1.3 V, respectively. The ideality factor $n$ can be calculated as a function of voltage by[25]

$$n = \frac{q}{2.3kT} \frac{1}{d(\log I)/dV} \quad (2)$$

where $q$ is electron charge, $k$ is the Boltzmann constant, $T$ is temperature, and $I$ is the current. At low bias, $n$ was 1.34 and 1.55 for the ($\bar{2}$01) SBD and the (010) SBD, respectively. Figure 4(b) presents current density and differential specific $R_{on}$ of the SBDs as a function of voltage in semi-log scale. Both SBDs showed a high on-current of ~ 1.3 kA/cm$^2$ and on/off ratio of ~ 10$^9$, which are among the highest values reported in vertical $\beta$-Ga$_2$O$_3$ SBDs.[3,6] At 1.3 kA/cm$^2$, $R_{on}$ was 0.56 and 0.77 m$\Omega\cdot$cm$^2$ for ($\bar{2}$01) and (010) SBDs, respectively. Figure 4(c) shows reported $R_{on}$ of $\beta$-Ga$_2$O$_3$ SBDs on the major crystal orientations, where both SBDs in this work demonstrated low $R_{on}$ values. The $\mu$ of the SBDs can be extracted by $t/qN_DR_{on}$ where $t$ is the substrate thickness and $N_D$ is the carrier concentration, considering a negligibly small contact resistance.[25] Taking the current spreading effect into consideration,[31] the electron mobility was calculated to be 125 cm$^2$/(V$\cdot$s) for the ($\bar{2}$01) SBD and 65 cm$^2$/(V$\cdot$s) for the (010) SBD, which are comparable to previous reports.[7,9,13] The difference in the electron mobilities of the two devices is due to anisotropic electron transport properties of $\beta$-Ga$_2$O$_3$.[13,14]

The I–V characteristics of the SBDs can be described by the thermionic emission model[10]

$$J = A^*T^2 \exp(-q\Phi_B/kT)\exp(qV/nkT - 1) \quad (3)$$

where $J$ is the current density, $A^*$ is the Richardson constant and $\Phi_B$ is the SBH. $A^*$ was calculated to be 41.1 A/(cm$^2$ K$^2$) using an effective electron mass of 0.34 m$_0$.[3] The extracted SBH were 1.05 eV for the ($\bar{2}$01) SBD and 1.20 eV for the (010) SBD. Figure 4(d) shows that (010) SBDs generally have higher SBH than ($\bar{2}$01) SBDs, which is consistent with previous reports,[33] and the larger $\Phi_{surf}$ of the (010) surface according to the XPS results. It's already been shown that the SBH of $\beta$-Ga$_2$O$_3$ SBDs is more dominated by the surface states and defects than by the actual metal used.[27,33] As shown in Figs. 2 and 3, the ($\bar{2}$01) and (010) surfaces have distinct Fermi level pinning and band bending. This indicates the interface states and defects between metal/$\beta$-Ga$_2$O$_3$ are different for ($\bar{2}$01) and (010) SBDs, leading to the discrepancy in the SBH between the two devices.

Figure 5 shows the C–V and 1/C$^2$–V characteristics of the ($\bar{2}$01) and (010) SBDs at a frequency of 1 MHz and RT. The built-in voltage $V_{bi}$ can be extracted from the intercept of 1/C$^2$ vs. V by[10,34]

$$1/C^2 = \frac{2}{q\varepsilon_0\varepsilon_r N_D}(V_{bi} - V - kT/q) \quad (4)$$

where $\varepsilon_0$ is the permittivity of the vacuum. The $V_{bi}$ of the ($\bar{2}$01) SBD was 1.41 V and that of the (010) SBD was 1.44 V. As depicted in the inset of Fig. 5(b), the Schottky barrier height can be expressed as[10]

$$q\Phi_B = qV_{bi} - q\Phi_{IL} + (E_C - E_F) \quad (5)$$

where $\Phi_{IL}$ is the image-force induced barrier height lowering, $E_C$ is the conduction band minimum, and $E_F$ is the Fermi level. $\Phi_{IL}$ is given by[10]

$$q\Phi_{IL} = \sqrt{qE_{SBD}/(4\pi\varepsilon_0\varepsilon_r)} \quad (6)$$

$$E_{SBD} = \sqrt{2qN_D V_{bi}/(\varepsilon_0\varepsilon_r)} \quad (7)$$

where $E_{SBD}$ is the electric field at the metal/semiconductor interface. ($E_C$–$E_F$) is calculated by $kT\ln(N_C/N_D)$ where $N_C$ is the effective density states. The $N_D$ was obtained from the slope of Eq. 4: $4.2 \times 10^{18}$ cm$^{-3}$ for the ($\bar{2}$01) SBD and $4.3 \times 10^{18}$ cm$^{-3}$ for the (010) SBD. After plugging in all the terms into Eq. 5, the SBH was 1.27 eV for ($\bar{2}$01) SBD and 1.30 eV for (010) SBD. $\Phi_B$ obtained from I-V are smaller than those from C-V for both devices. This is usually attributed to the spatially inhomogeneous SBH caused by the interfacial states and defects between metal/semiconductor.[33,35-37] Furthermore, the $\Phi_B$ of two SBDs from the C-V data have only a small difference of 0.03 eV compared with 0.15eV $\Phi_B$ difference from the I-V data. The C-V SBH is more influenced by the doping concentrations of the semiconductors and doesn't involve current conduction, while the I-V SBH represents the barrier height for current flow.[33] Therefore, the C-V SBH is insensitive to the crystal orientations and surface properties of $\beta$-Ga$_2$O$_3$ that dominate the current conduction.

Figures 6(a) and 6(b) present the temperature-dependent I–V characteristics for the ($\bar{2}$01) and (010) SBDs. Based on Eq. 3, the SBH and ideality factor of the devices were extracted as a function of temperature shown in Fig. 6(c). For the ($\bar{2}$01) SBD, $\Phi_B$ increased from 1.05 eV to 1.18 eV and $n$ decreased from 1.34 to 1.20 with increasing temperature. For the (010) SBD, $\varPhi_B$ increased from 1.20 eV to 1.36 eV and $n$ decreased from 1.55 to 1.29 with increasing temperature. The temperature dependence of the ideality factor, also called "$T_0$ anomaly", is caused by the spatial inhomogeneity of SBH as a result of surface states and defects at the metal/semiconductor interface,[36] as confirmed by the previous XPS results. Similar phenomena have also been observed in other semiconductors.[36,37] The ideality factor can be described as a function of temperature by[36]

$$n = 1 + T_0/T \tag{8}$$

where $T_0$ is a constant associated with the standard deviation of the SBH distribution. Fitting experimental data with Eq. 8 yielded a $T_0$ of 88 K for the ($\bar{2}$01) SBD and a $T_0$ of 133 K for the (010) SBD. The smaller $T_0$ of the ($\bar{2}$01) SBD indicates a more uniform SBH distribution due to different surface properties. In Fig. 6(d), there was a well-known linear relationship between the SBH and ideality factor for both devices due to the inhomogeneous Schottky barrier interfaces.[34,36] By extrapolation, the homogenous SBH when $n = 1$ was 1.33 eV for the ($\bar{2}$01) SBD and 1.53 eV for the (010) SBD, which are larger than their inhomogeneous SBH ( 1.06 eV and 1.20 eV, respectively). The device performance metrics of the two devices from forward I-V and C-V characteristics are summarized in Table I.

Figure 7 shows the experimental and theoretical data of temperature-dependent reverse leakage current of the two SBDs at – 6 V. The reverse leakage current of SBDs above RT is usually characterized by two models.[38] The first model is the two-step trap-assisted tunneling mechanism, where the electrons in the metal first are thermally excited to the trap states and then tunnel through the Schottky barrier [shown the inset of Fig. 7(a)]. The reverse leakage current is proportional to $exp(-E_A/kT)$ where $E_A$ is the activation energy. A good agreement was obtained between the experiment and this model in the Arrhenius plot in Fig. 7(a). $E_A$ was 42 meV for the ($\bar{2}$01) SBD and 71 meV for the (010) SBD. Another possible model is the one-dimensional variable-range-hopping conduction (1D-VRH) model, where the electrons in the metal first fall into defect states associated with a dislocation near or below the Fermi level and are then transported into the semiconductor by hopping conduction [shown the inset of Fig. 7(b)]. In this model, the conductivity is given by[38]

$$\sigma = \sigma_0 \exp[-(T_1/T)^{1/2}] \tag{9}$$

where $\sigma_0$ is a constant and $T_1$ is the characteristic temperature. Figure 7(b) shows a good linear fitting for the log of the conductivity as a function of $1/T^{1/2}$ between experimental and simulation data. It is not clear yet which model is the dominant mechanism based on current data and further investigations are undergoing. In addition, the ($\bar{2}$01) SBD exhibited higher reserve leakage current than the (010) SBD, which can be attributed to lower SBH and smaller $E_A$ of the ($\bar{2}$01) SBD.

In summary, vertical $\beta$-Ga$_2$O$_3$ SBDs were fabricated on single-crystal ($\bar{2}$01) and (010) substrates grown by EFG method, followed by comprehensive device analysis. The devices showed excellent forward characteristics with a record low $R_{on}$, a high on-current, and a high electron mobility. The ($\bar{2}$01) SBD showed smaller $V_{on}$ and SBH than the (010) SBD, attributed to anisotropic crystal structure and surface properties, as confirmed by XPS results. Temperature-dependent I-V characteristics revealed the inhomogeneous nature of the SBH for both devices, where the ($\bar{2}$01) SBD exhibited a more uniformly distributed SBH. The reverse leakage current of both devices were simulated by the two-step trap-assisted tunneling model and the 1D-VRH model. Good agreements between experimental and theoretical data were obtained for both models. Further investigation is demanded to determine the dominant mechanism. This work shows crystalline anisotropy of $\beta$-Ga$_2$O$_3$ can impact the electrical properties of vertical SBDs, and possibly transistors as well. Special attention needs to be paid to this anisotropic crystal structure when designing $\beta$-Ga$_2$O$_3$ electronics.

**Acknowledgements**

We gratefully acknowledge the use of facilities within the LeRoy Eyring Center for Solid State Science at Arizona State University. The samples were provided by Tamura Corporation. The device fabrication was done in the Center for Solid State Electronics Research at Arizona State University.

**Figure captions**

Figure 1 Rocking curves by HRXRD for (a) ($\bar{2}$01) and (b) (010) $\beta$-Ga$_2$O$_3$ substrates. Representative AFM images of the substrates in (c) two dimension and (d) three dimension.

Figure 2 (a) Unit cell of $\beta$-Ga$_2$O$_3$ crystal with ($\bar{2}$01) and (010) planes labeled. The atomic structures of (b) ($\bar{2}$01) plane and (c) (010) plane.

Figure 3 XPS valence band spectra of (a) ($\bar{2}$01) and (b) (010) $\beta$-Ga$_2$O$_3$. $E_{VBM}$ were also extracted by extrapolation. The insets indicate the upward band bending at the surfaces.

Figure 4 (a) Current and ideality factor as a function of forward bias in linear scale. The $V_{on}$ were also obtained by linear extrapolation. (b) Current density and $R_{on}$ versus forward bias in semi-log scale. (c) Comparison of $R_{on}$ of reported $\beta$-Ga$_2$O$_3$ SBDs on different orientations. (d) Comparison of Schottky barrier height of ($\bar{2}$01) and (010) $\beta$-Ga$_2$O$_3$ SBDs.

Figure 5 (a) C–V and (b) 1/C$^2$–V characteristics of ($\bar{2}$01) and (010) SBDs at 1 MHz. The inset in the right figure shows the band diagram of Pt/$\beta$-Ga$_2$O$_3$ Schottky interface.

Figure 6 The temperature-dependent forward I–V characteristics for (a) the ($\bar{2}$01) SBD and (b) the (010) SBD. (c) Ideality factor and Schottky barrier height as a function of temperature for the two devices. (d) Schottky barrier height vs. ideality factor.

Figure 7 (a) Arrhenius plot of reverse leakage current of the two SBDs with activation energy extracted. The inset shows the electron transport in two-step trap-assisted tunneling model. (b) Conductivity as a function of $1/T^{1/2}$ for the two SBDs. 1D-VRH conduction model (dot line) is used to fit the data. The inset shows the electron transport in this model.

**Figures**

**Figure 1**

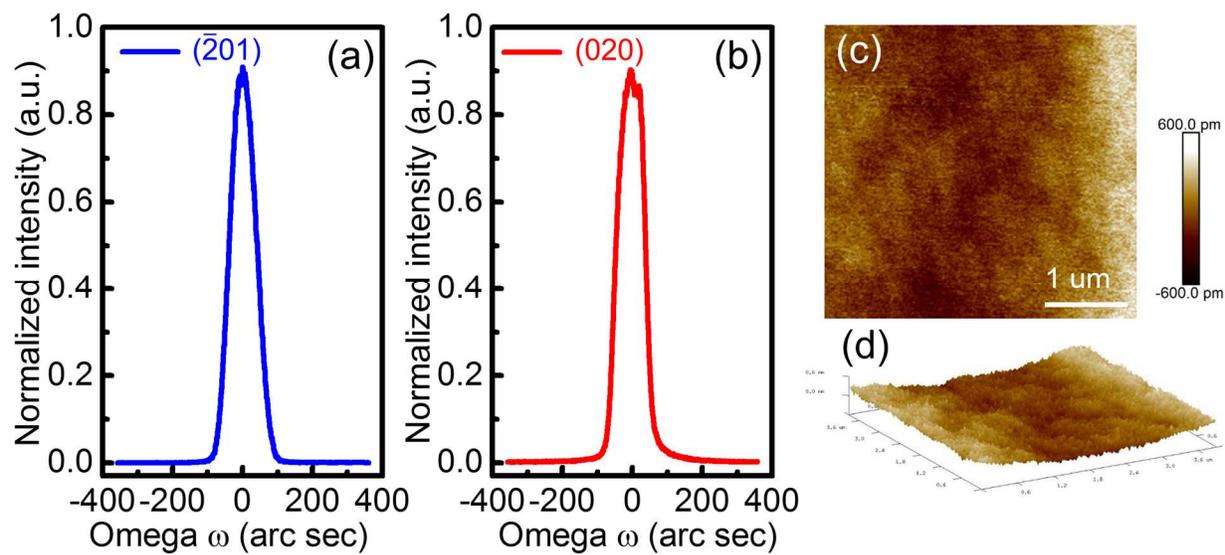

**Figure 2**

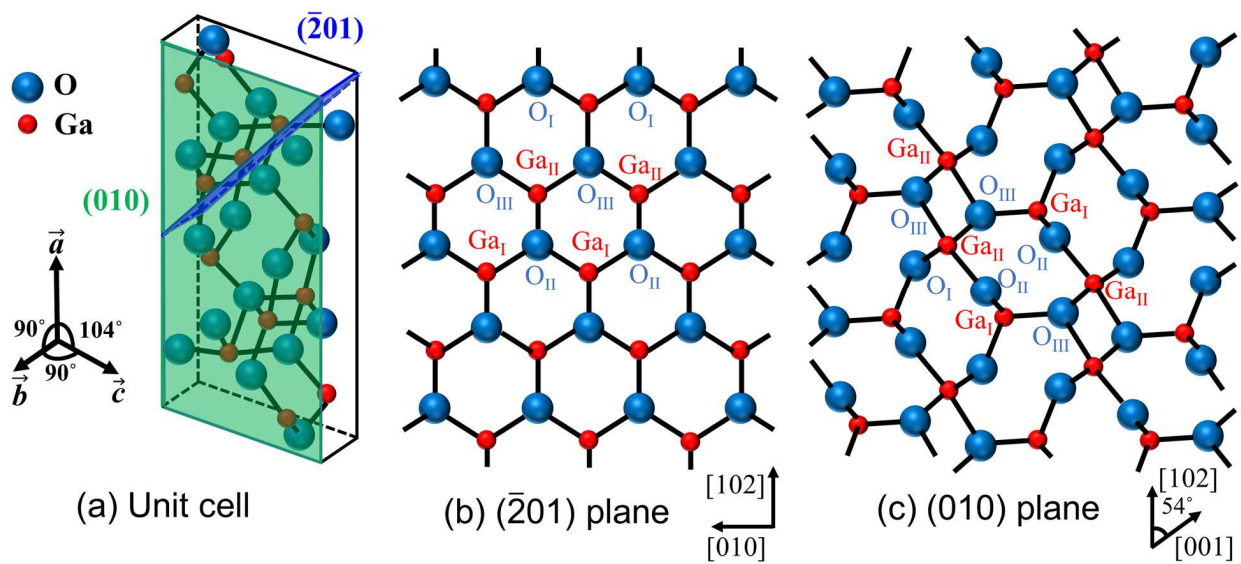

**Figure 3**

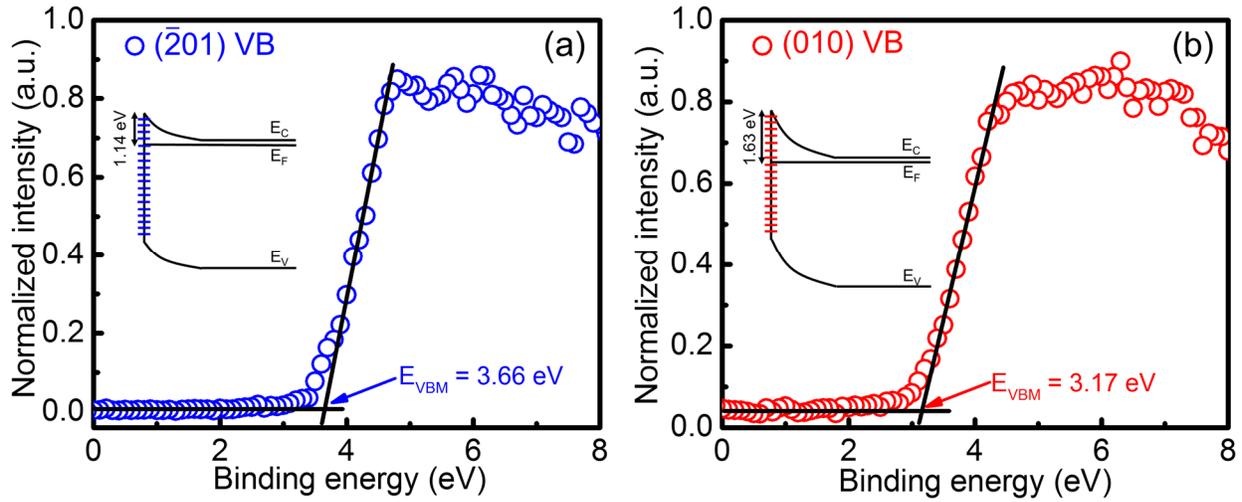

**Figure 4**

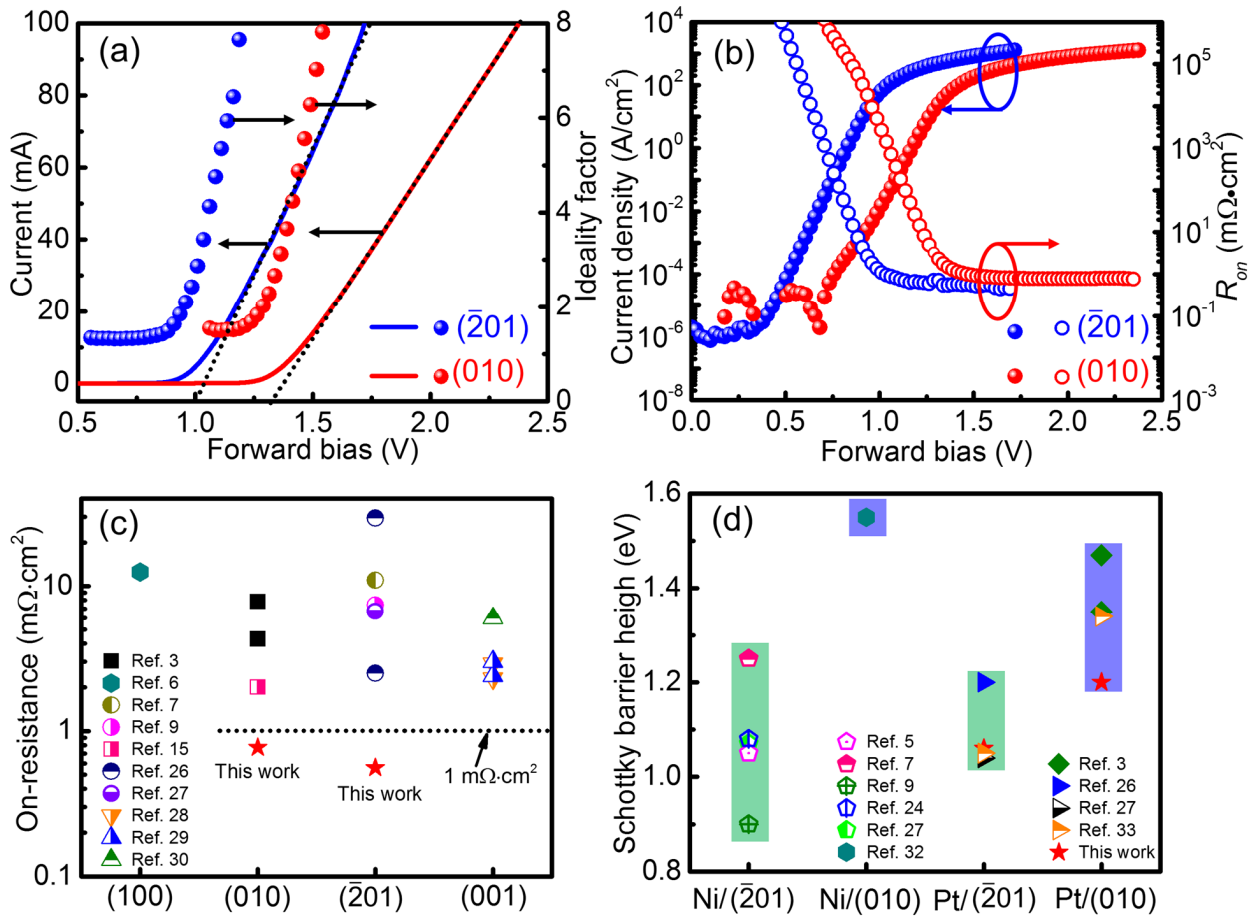

**Figure 5**

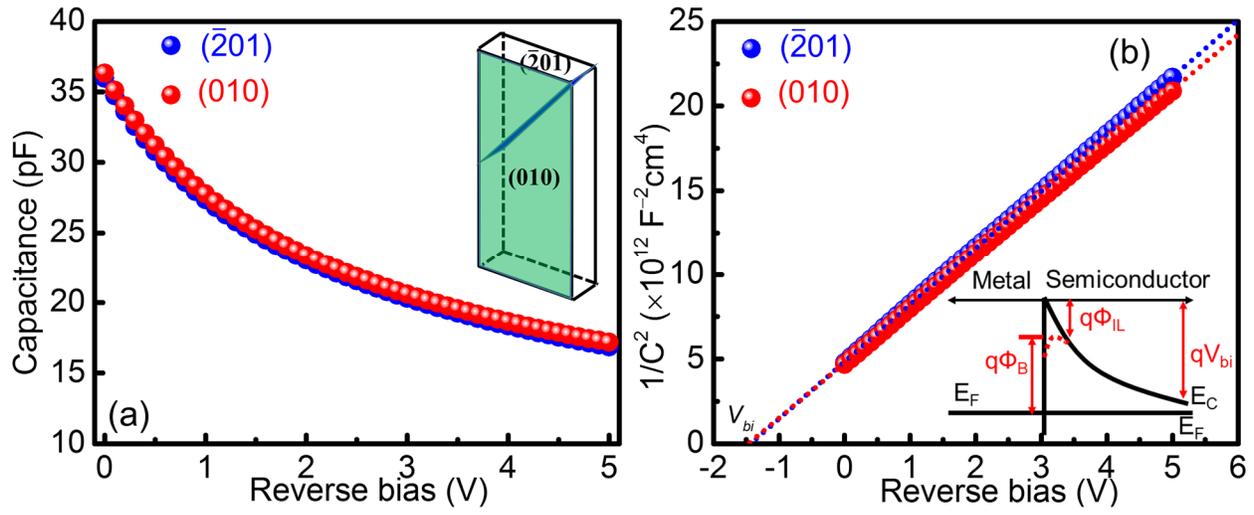

**Figure 6**

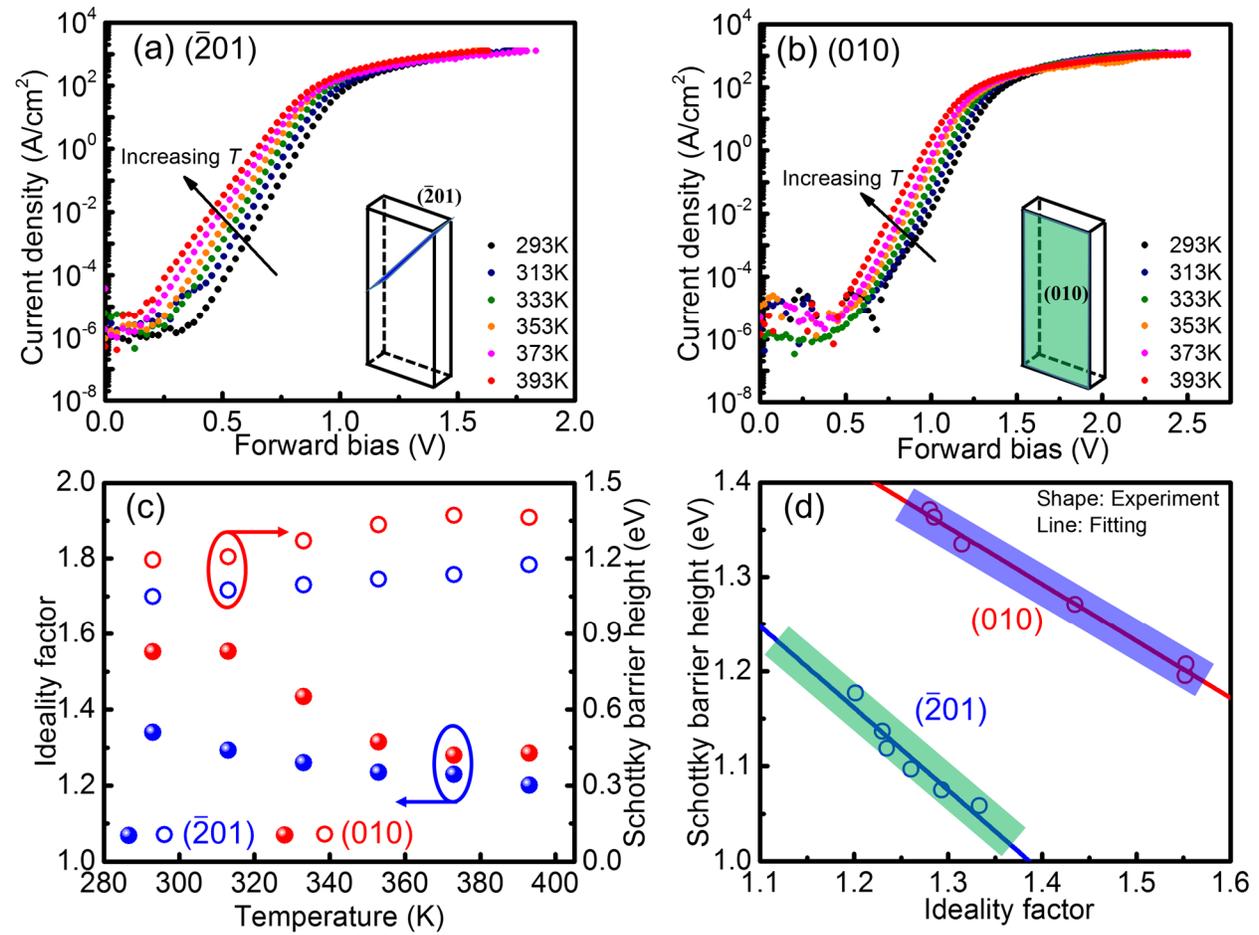

**Figure 7**

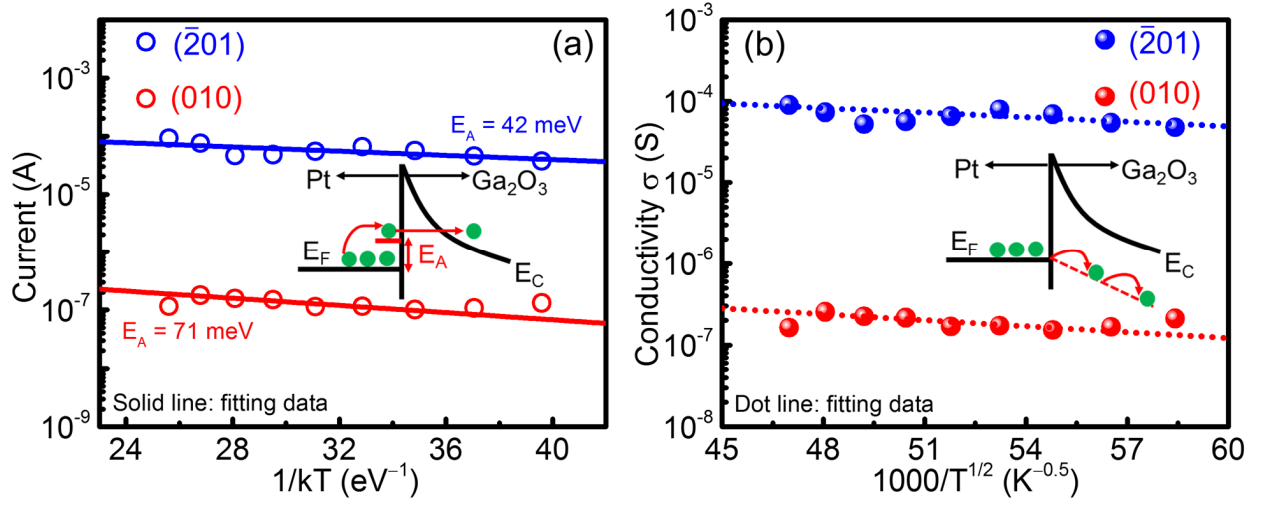

**Tables**

TABLE I. Summary of device parameters for the ($\bar{2}$01) and (010) SBDs.

| Sample | $R_{on}$ (mΩ·cm$^2$) | $V_{on}$ (V) | $n$ | Mobility [cm$^2$/(V·s)] | $\Phi_{B, I-V,}$ inhomogeneous (eV) | $\Phi_{B, I-V,}$ homogeneous (eV) | $V_{bi}$ (V) | $\Phi_{B, C-V}$ (eV) |
|---|---|---|---|---|---|---|---|---|
| ($\bar{2}$01) | 0.56 | 1.0 | 1.34 | 125 | 1.05 | 1.33 | 1.41 | 1.27 |
| (010) | 0.77 | 1.3 | 1.55 | 65 | 1.20 | 1.53 | 1.44 | 1.30 |